\documentclass[prb,twocolumn,showpacs,longbibliography,superscriptaddress]{revtex4-1}

\usepackage{graphicx}

\usepackage{amsmath}
\usepackage{amssymb}
\usepackage{dcolumn}
\usepackage{bm}
\usepackage{wasysym}
\usepackage{dsfont}
\usepackage{enumitem}
\usepackage{hyperref}
\usepackage{braket}
\usepackage[most]{tcolorbox}

\newcommand{\ii}{{\rm i}}

\def\cC{{\cal C}}

\def\cZ{{\cal Z}}
\def\cN{{\cal N}}
\def\U{{\rm U}}
\def\errchn{\{\varepsilon_\mu\}}
\newcommand{\Zb}{ \mathbb{Z} }
\newcommand{\Tr}{{\rm Tr}}
\newcommand{\dtilde}[1]{\tilde{\raisebox{0pt}[0.85\height]{$\tilde{#1}$}}}

\usepackage{xcolor}

\begin{document}

\title{A Mechanism for the R\'enyi Hierarchy of Decoherence-Induced Phase Transitions}
\author{Zhou Yang}
\affiliation{Department of Physics, Cornell University, Ithaca, New York 14853, USA}

\author{Yuri D. Lensky}
\affiliation{Google Quantum AI}

\author{Chao-Ming Jian}
\affiliation{Department of Physics, Cornell University, Ithaca, New York 14853, USA}

\begin{abstract}
Decoherence-induced phase transitions (DIPTs) are associated with singular changes in a quantum system's entanglement structure as the decoherence strength varies. In many cases, their critical strengths depend monotonically on the R\'enyi index of the entanglement diagnostic, forming a R\'enyi hierarchy. We identify a general mechanism for this hierarchy based on correlation inequalities in replicated statistical models that describe these DIPTs. We demonstrate it in $\Zb_2$ topological stabilizer codes subject to independent phase-flip decoherence on each qubit and in a decohered rotor model with strong $\U(1)$ symmetry. When the relevant correlations diagnose the DIPTs, these inequalities imply that the critical decoherence strength is nondecreasing with integer R\'enyi index $R\geq2$. We show that this mechanism also applies to a class of models with correlated decoherence.
\end{abstract}

\date{\today}

\maketitle

\section{Introduction}
Decoherence-induced phase transitions (DIPTs) describe singular changes in the correlation and entanglement structure of out-of-equilibrium quantum systems as the strength of decoherence varies. They have attracted growing interest because they provide new perspectives and tools for understanding decoding thresholds in quantum error-correction codes~\cite{dennis2002topological,Preskill_3dToricCodeThreshold,Flammia2021,FanBaoTopoMemory,BaoFanError_Field_Double,LeeJianXu2023,SangZouHsieh2024RG, Mong2025_ReplicaTopo, SuYangJian2024, Martin-Delgado_FractonThreshold, Wang2025SelfDualCriticality, Lee2025ExactCoherentInformation, NiwaLee2025CSSCoherentInformation,Colmenarez2024CoherentThresholds,Lyons2024StabilizerDecoherence,BehrendsBeri2025BeyondPauli} and help extend familiar concepts in equilibrium quantum systems, such as spontaneous symmetry breaking and topological order, to nonequilibrium mixed states~\cite{Lessa2025_SWSSB,LeeYouXu2022,LeeJianXu2023,FanBaoTopoMemory,BaoFanError_Field_Double,EllisonCheng2025,SohalPrem2025,Mong2025_ReplicaTopo,BowenLee2025MixedTOAxiom,Wang2025MixedStateTO,Zhang2025HigherFormSWSSB,Hauser2026SWSSBHydro,XuBloch2026ExpSWSSB,You2024PurificationSWSSB,HuangLucas2025,MaWang2025ASPT,WangEhud2025NoisyFQH,Grover2024Separability,SangZouHsieh2024RG,Lee2026Scrambling,OgunnaikeLee2023Lind}. 

DIPTs can often be diagnosed using information-theoretic quantities, such as the von Neumann entropy and coherent information, and their R\'enyi generalizations. For a given decohered quantum system, singularities in a chosen R\'enyi diagnostic, such as the R\'enyi entropy $S_R$, can occur at different critical decoherence strengths for different R\'enyi indices $R$. In many studied examples, the critical strength depends monotonically on $R$.~\cite{Sommers2026Haar,SuYangJian2024,FanBaoTopoMemory,KimAltmanLee2026SYKThreshold} We refer to this interesting behavior as the R\'enyi hierarchy of DIPTs. 

Beyond its implications for many-body entanglement, this hierarchy has an operational interpretation in quantum error correction. In decohered topological stabilizer codes such as the noisy toric code, the critical decoherence strength for the DIPT in the von Neumann entropy $S_{R\rightarrow1}$ corresponds to the optimal decoding threshold without postselection~\cite{dennis2002topological,BaoFanError_Field_Double,FanBaoTopoMemory,LeeJianXu2023}, whereas transitions at integer R\'enyi indices $R\geq2$ correspond to thresholds of decoding protocols based on postselection across $R$ copies of the decohered code~\cite{Mong2025_ReplicaTopo}. The R\'enyi hierarchy thus translates into a hierarchy of decoding thresholds, further motivating the search for mechanisms that enforce it.

In this work, we identify a general mechanism for this hierarchy in a class of decohered systems, including $\mathbb{Z}_2$ topological stabilizer codes under Pauli noise and a decohered rotor system with $\U(1)$ charge conservation. In these settings, DIPTs in the R\'enyi entropy $S_R$ admit descriptions in terms of replicated statistical models with random couplings, generalizing the random-bond Ising model associated with the noisy toric code~\cite{dennis2002topological,Flammia2021,Preskill_3dToricCodeThreshold} and its replicated counterparts~\cite{FanBaoTopoMemory,BaoFanError_Field_Double,LeeJianXu2023}. We show that, under suitable positivity conditions, analogues of the Griffiths--Kelly--Sherman (GKS) inequalities\cite{Griffiths1967,KellySherman1968} for ferromagnets and their Ginibre generalizations\cite{ginibre1970general} establish an ordering of correlation functions in the replicated statistical models for R\'enyi index $R\geq2$. When the asymptotic behavior of these correlations diagnoses the DIPTs, this ordering implies the R\'enyi hierarchy of critical decoherence strengths of the associated DIPTs. We also discuss conditions under which the hierarchy may fail.

The rest of the paper is organized as follows. In Sec.~\ref{sec:Z2}, we establish the GKS-based mechanism for $\mathbb{Z}_2$ CSS topological stabilizer codes decohered by independent phase-flip errors and discuss extensions to more general Pauli noise. Sec.~\ref{sec:Z2extended} extends the discussion to correlated phase-flip errors. In Sec.~\ref{sec:U1}, we establish the R\'enyi hierarchy in a decohered rotor model with strong $\U(1)$ symmetry which is motivated by charge-informed error correction and $\U(1)$ strong-to-weak spontaneous symmetry breaking.

\section{R\'enyi hierarchy in $\Zb_2$ topological stabilizer codes}
\label{sec:Z2}
In this section, we show how inequalities analogous to the GKS inequalities for ferromagnets \cite{Griffiths1967,KellySherman1968,ginibre1970general} provide a general mechanism for the R\'enyi hierarchy in the DIPTs in $\Zb_2$ topological stabilizer codes. We focus on Calderbank--Shor--Steane (CSS) codes under phase-flip decoherence to illustrate the GKS-based mechanism for the R\'enyi hierarchy. At the end of this section, we discuss possible extensions to non-CSS codes and more general Pauli noise.

Consider a translationally invariant $\Zb_2$ topological CSS stabilizer code $\cC$ on a lattice of qubits labeled by $\mu$. We assume that the number of logical qubits remains finite as the system size increases, while the minimum weight of a nontrivial logical operator diverges. For simplicity, we consider the decoherence caused by independent stochastic phase-flip errors with a uniform probability $p$ on each qubit. The corresponding decoherence channel is
\begin{align}
    {\cal N}_{\rm tot} &= \prod_\mu {\cal N}_\mu,
    \nonumber\\
   {\rm with}~~~ {\cal N}_\mu(\rho) &= (1-p)\rho + p Z_\mu \rho Z_\mu,
    \label{eq:PhaseFlip}
\end{align}
where $Z_\mu$ is the Pauli-$Z$ operator on qubit $\mu$. We restrict to the range $0<p<1/2$, since the channel at $p>1/2$ is related to that at $1-p$ by conjugation with a global unitary action $\prod_\mu Z_\mu$, which preserves the nature of the DIPTs. $p$ controls the strength of decoherence and will also be referred to as the error rate.

The code's initial state $\rho_0$ is a pure logical state satisfying all stabilizer constraints. After decoherence, its $R$th R\'enyi entropy is
\begin{align}
    S_R = -\frac{1}{R-1}
    \log \Tr\left({\cal N}_{\rm tot}(\rho_0)^R\right).
\end{align}
As $p$ increases, a DIPT can occur at an $R$-dependent critical value $p_c(R)$, where $S_R$ becomes nonanalytic in the thermodynamic limit. As $R\rightarrow1$, $S_R$ approaches the von Neumann entropy. The associated DIPT is identified with the code's optimal decoding threshold \cite{dennis2002topological} and the fundamental change in mixed-state topological order \cite{FanBaoTopoMemory,BaoFanError_Field_Double,LeeJianXu2023,Grover2024Separability,SangZouHsieh2024RG, EllisonCheng2025, Wang2025MixedStateTO,SohalPrem2025,Zhang2025HigherFormSWSSB}. For integer R\'enyi index $R\geq2$, Ref.~\onlinecite{Mong2025_ReplicaTopo} relates the DIPT to the error threshold of a postselection-based error-correction protocol involving $R$ copies of the decohered code. 

In many previously studied examples of topological stabilizer codes (see, e.g., Refs.~\onlinecite{FanBaoTopoMemory,BaoFanError_Field_Double,SuYangJian2024}), $p_c(R)$ increases with the R\'enyi index $R$. This behavior is the R\'enyi hierarchy of DIPTs we investigate here. In fact, these examples extend beyond CSS codes and/or models with only phase-flip decoherence. In the following, we still largely focus on the R\'enyi hierarchy of DIPTs in CSS codes under phase-flip decoherence for concreteness.

The DIPTs in the decohered topological stabilizer codes admit a statistical-mechanical description \cite{dennis2002topological,Flammia2021,FanBaoTopoMemory,LeeJianXu2023}. For the channel in Eq.~\eqref{eq:PhaseFlip}, an error chain $E$ is the set of qubits $\mu$ on which a phase flip occurs. We represent it by Ising variables
\begin{align}
    \varepsilon_\mu =
    \begin{cases}
        -1, & \mu\in E,\\
        +1, & \mu\notin E.
    \end{cases}
\end{align}
The probability of this error chain is
\begin{align}
    P_{\{\varepsilon_\mu\}}
    = p^{N/2}(1-p)^{N/2}
    \exp\left\{J\sum_\mu\varepsilon_\mu\right\},
    \label{eq:ErrorChainProb}
\end{align}
where $J=\frac{1}{2}\log\frac{1-p}{p} > 0$ and $N$ is the total number of qubits.

Phase-flip errors violate only the $X$-type stabilizers of the CSS code $\cC$. Error chains with the same syndrome form an equivalence class and differ by products of $Z$-type stabilizers and, potentially, logical operators. Since the number of logical qubits is finite, distinctions between logical sectors contribute only subextensively to $S_R$ and do not affect our subsequent analysis of its extensive part and the associated DIPT in the thermodynamic limit. To enumerate error chains for a fixed syndrome (and logical sector), we assign a classical Ising spin $\sigma_v=\pm1$ to each $Z$-type stabilizer generator labeled by $v$. For an error chain $E=\errchn$, the total probability of all the 
error chains within the same class is given by
\cite{dennis2002topological,Flammia2021}
\begin{align}
    \cZ_{\{\varepsilon_\mu\}}
    \propto p^{N/2}(1-p)^{N/2}
    \sum_{\{\sigma_v=\pm1\}}
     \exp\left\{
        J\sum_\mu\varepsilon_\mu
        \prod_{\mu\in v}\sigma_v
    \right\}.
    \label{eq:SyndromeProb}
\end{align}
Here, $\prod_{\mu\in v}\sigma_v$ denotes the product over the $Z$-type stabilizer generators $v$ whose support contains qubit $\mu$. The proportionality constant in Eq. \eqref{eq:SyndromeProb} depends on relations among the stabilizer generators, which are unimportant for our discussion. From a statistical-mechanics perspective, Eq.~\eqref{eq:SyndromeProb} is the partition function of an Ising-type statistical model with interactions $\prod_{\mu\in v}\sigma_v$ and random coupling signs $\varepsilon_\mu$. For the standard 2D toric code, it reduces to the 2D random-bond Ising model.\cite{dennis2002topological}

The decohered state $\cN_{\rm tot}(\rho_0)$ is block diagonal. Each block corresponds to a unique syndrome that appears with probability $ \cZ_{\{\varepsilon_\mu\}}$. Up to subextensive contributions from the logical qubits and unimportant normalization terms, the bulk singularities of $S_R$ are captured by $-\frac{1}{R-1}\log\cZ_R$, where, for $R\geq2$,
\begin{align}
    \cZ_R
    &= \sum_{\{\varepsilon_\mu=\pm1\}}
    \left(\cZ_{\{\varepsilon_\mu\}}\right)^R
    \nonumber\\
    &\propto
    \sum_{\{\varepsilon_\mu,\,\sigma_v^{(a)}\}}
    \exp\left\{
        J\sum_\mu\varepsilon_\mu
        \sum_{a=1}^{R}
        \prod_{\mu\in v}\sigma_v^{(a)}
    \right\}.
\end{align}
Here, $\sigma_v^{(a)}$ is the Ising spin in replica $a=1,\ldots,R$. Under the change of variables
$\varepsilon_\mu\rightarrow\varepsilon_\mu\prod_{\mu\in v}\sigma_v^{(R)}$
and
$\sigma_v^{(a)}\rightarrow\sigma_v^{(a)}\sigma_v^{(R)}$
for $a=1,\ldots,R-1$, the Ising spins $\sigma_v^{(R)}$ decouple and contribute only an overall factor. Thus,
\begin{align}
    \cZ_R&\propto
    \sum_{\{\varepsilon_\mu,\,\sigma_v^{(a)}\}} \exp\left\{
        J\sum_\mu\varepsilon_\mu
        \left(
            1+\sum_{a=1}^{R-1}
            \prod_{\mu\in v}\sigma_v^{(a)}
        \right)
    \right\},
    \label{eq:ZR_fixed}
\end{align}
which takes the form of a replicated Ising-type statistical model. For integer $R\geq2$, the error variables $\errchn$ are effectively annealed disorder coupled to $R-1$ Ising spin replicas. The replica limit $R\rightarrow1$ extracts quenched averages with the disorder distribution in Eq.~\eqref{eq:ErrorChainProb}. The same coupling $J$ controls both the spin interactions and the disorder distribution, which is the Nishimori condition.

We first outline the general intuition behind the DIPTs described by $\cZ_R$. At small error rate $p$, or equivalently large positive $J$, the replicated statistical model favors configurations that maximize every interaction term in the Boltzmann weight exponent. These include the fully polarized configuration $\varepsilon_\mu=\sigma_v^{(a)}=1$. A transition driven by increasing $p$, or decreasing $J$, can therefore be viewed as the loss of the ``ferromagnet-like order" associated with this configuration. For the 2D toric code, this picture is made precise: the DIPTs correspond to ferromagnetic transitions in $\cZ_R$ \cite{dennis2002topological,FanBaoTopoMemory,BaoFanError_Field_Double,SuYangJian2024,LeeJianXu2023}. For a general CSS code $\cC$, however, the large-$J$ phase need not be a conventional ferromagnet. For example, the 3D toric code maps to the 3D random-plaquette $\Zb_2$ gauge model and its replicated counterparts \cite{Preskill_3dToricCodeThreshold}. At sufficiently large $J$, these models exhibit a deconfined phase, in which $\varepsilon_\mu=\sigma_v^{(a)}=1$ is one of the favored configurations.

Although the nature of the large-$J$ phase depends on the code $\cC$, we expect the DIPT out of this phase to be diagnosed by order parameters constructed from suitable products of Ising spins. For appropriately chosen order parameters, their expectation values or correlations that characterize the order should be parametrically larger in the large-$J$ phase than in the small-$J$ phase.  This expectation is based on the fact that the configuration with $\varepsilon_\mu=\sigma_v^{(a)}=1$ is always favorable in the large-$J$ limit. Also, it is consistent with many examples of decohered codes studied before\cite{dennis2002topological,FanBaoTopoMemory,BaoFanError_Field_Double,SuYangJian2024,LeeJianXu2023, Preskill_3dToricCodeThreshold,Matin-DelgadoColorCodes,Kubica2018ColorCodeThresholds, Martin-Delgado_FractonThreshold, canossa2023exotic}. We will show that the expectation values of arbitrary Ising spin products are nondecreasing with $R$ for $R\geq2$ at fixed $J$, or equivalently fixed $p$. When these observables diagnose the DIPT, this ordering implies that $p_c(R)$ is also nondecreasing with $R$.

To this end, we define the Ising spin product
$\sigma_V^{(A)}\equiv\prod_{v\in V,\,a\in A}\sigma_v^{(a)}$,
where $V$ is a set of Ising spin locations and $A$ is a set of replica indices. We consider its expectation value $\langle\sigma_V^{(A)}\rangle_R$ in the replicated statistical model $\cZ_R$ [Eq. \eqref{eq:ZR_fixed}]. This notation implicitly assumes $A\subseteq\{1,\ldots, R-1\}$, which is necessary for $\langle\sigma_V^{(A)}\rangle_R$ to be well defined (in the representation Eq. \eqref{eq:ZR_fixed} of $\cZ_R$). Of particular interest are products $\sigma_V^{(A)}$ that characterize the relevant order. For the 2D toric code, choosing $\sigma_V^{(A)}=\sigma_v^{(1)}\sigma_{v'}^{(1)}$ gives the two-point function of the Ising order parameter. For the 3D toric code with the conventional assignment of $X$-type star and $Z$-type plaquette stabilizers, the phase-flip problem maps to the 3D random-plaquette $\mathbb{Z}_2$ gauge model. Taking $V$ to be a closed loop on the dual lattice and $A=\{1\}$ gives a Wilson loop, whose large-loop behavior distinguishes confinement from deconfinement.

For $J>0$ and integers $R>R'\geq2$, we will prove an ordering of correlations
\begin{align}
    \langle\sigma_V^{(A)}\rangle_R
    \geq
    \langle\sigma_V^{(A)}\rangle_{R'}.
    \label{eq:GKS}
\end{align}
Thus, at fixed $J>0$, these correlations are nondecreasing with $R$. If their long-distance or large-loop behavior characterizes the DIPT out of the large-$J$ phase, then, as $J$ decreases from $+\infty$, the large-$J$ phase in $\cZ_R$ persists down to a critical coupling no larger than that for $\cZ_{R'}$. Since $J=\frac{1}{2}\log\frac{1-p}{p}$ decreases with $p$, this implies that $p_c(R)$ is nondecreasing with $R$.

Eq.~\eqref{eq:GKS} establishes this ordering only for integer $R\geq2$. We expect it to extend to the limit $R\rightarrow1$, where $\langle\sigma_V^{(A)}\rangle_{R \rightarrow 1}$ represents the expectation value of $\sigma_V^{(A)}$ with quenched disorder $\{\varepsilon_\mu\}$ and $p_c(R\rightarrow1)$ corresponds to the optimal decoding threshold of the code $\cC$. Establishing this extension requires a separate comparison with the quenched expectation value $\langle\sigma_V^{(A)}\rangle_{R\rightarrow1}$. We leave this question to future work.

We now prove Eq.~\eqref{eq:GKS} using an argument analogous to the standard GKS proof. Since $\cZ_R$ and $\cZ_{R'}$ are positive, it suffices to show that
$\cZ_R\cZ_{R'}(\langle\sigma_V^{(A)}\rangle_R-\langle\sigma_V^{(A)}\rangle_{R'})\geq0$.
Suppressing an overall positive normalization factor, we write
\begin{widetext}
   \begin{align}
    & \cZ_R \cZ_{R'}  \left(\langle \sigma_V^{(A)} \rangle_R - \langle \sigma_V^{(A)} \rangle_{R'}\right) 
    \nonumber \\
    &= \sum_{\{\varepsilon_\mu,\,  \sigma_v^{(a)} \}}  \sum_{\{\tilde \varepsilon_\mu,\,  \tilde\sigma_v^{(a)} \}} \left( \sigma_V^{(A)} - \tilde \sigma_V^{(A)} \right) \exp\left\{ J \sum_\mu  \varepsilon_\mu \left(1+\sum_{a=1}^{R-1} \prod_{\mu \in v} \sigma_v^{(a)} \right) +  J \sum_\mu  \tilde\varepsilon_\mu \left(1+\sum_{a=1}^{R'-1} \prod_{\mu \in v} \tilde \sigma_v^{(a)} \right) \right\}. 
    \nonumber \\
    &= \sum_{\{\dtilde \varepsilon_\mu,\,  \dtilde \sigma_v^{(a)} \}}  
    \left( 1 - \dtilde{\sigma}_V^{(A)}  \right) 
    \nonumber \\
    & ~~~~~  \times \left[ \sum_{\{\varepsilon_\mu,\,  \sigma_v^{(a)} \}} 
    \sigma_V^{(A)}  
    \exp\left\{ J \sum_\mu  \varepsilon_\mu 
    \sum_{a=R'}^{R-1} \prod_{\mu \in v} \sigma_v^{(a)}  
    + J \sum_\mu   \varepsilon_\mu \left( \left(1+\dtilde \varepsilon_\mu \right)+\sum_{a=1}^{R'-1} \left( 1+ \dtilde \varepsilon_\mu \prod_{\mu \in v} \dtilde \sigma_v^{(a)} \right) \prod_{\mu \in v}   \sigma_v^{(a)} \right) \right\} \right] \geq 0 . 
\end{align} 
\end{widetext}
The first equality follows from expanding the two expectation values as statistical sums. The untilded variables, $\sigma_v^{(a)}$ and $\varepsilon_\mu$, belong to the $\cZ_R$ model, while the tilded variables, $\tilde\sigma_v^{(a)}$ and $\tilde\varepsilon_\mu$, belong to the $\cZ_{R'}$ model. For the second equality, we substitute the tilded variables using new Ising variables
$\dtilde{\varepsilon}_\mu=\tilde\varepsilon_\mu\varepsilon_\mu$
and
$\dtilde{\sigma}_v^{(a)}=\tilde\sigma_v^{(a)}\sigma_v^{(a)}$
for $a=1,\ldots,R'-1$, and define
$\dtilde{\sigma}_V^{(A)}=\prod_{v\in V,\,a\in A}\dtilde{\sigma}_v^{(a)}$. Note that every Ising variable squares to one.
Reorganizing the exponent and interchanging the sums then yields the expression above.

For each fixed configuration of the variables $\dtilde{\varepsilon}_\mu$ and $\dtilde{\sigma}_v^{(a)}$, the  square bracket is an unnormalized expectation of $\sigma_V^{(A)}$ in a ``renormalized" Ising-type model whose dynamical spins include both $\varepsilon_\mu$ and $\sigma_v^{(a)}$. All its renormalized couplings are nonnegative: $J>0$ is positive, while the factors $(1+\dtilde{\varepsilon}_\mu)$ and $\left(1+\dtilde{\varepsilon}_\mu\prod_{\mu\in v}\dtilde{\sigma}_v^{(a)}\right)$ are non-negative. The first GKS inequality for nonnegative multispin couplings guarantees that the expectation value of $\sigma_V^{(A)}$ in the renormalized model is nonnegative. Since $1-\dtilde{\sigma}_V^{(A)}$ is also nonnegative, the entire sum is nonnegative. This proves Eq.~\eqref{eq:GKS}.

For $R\geq3$, suppose an order parameter of interest in $\cZ_R$ is an Ising spin product involving all $R-1$ replicas. We partition the replica indices into two disjoint sets $A$ and $B$, with $1\leq |A|,|B|\leq R-2$ and $|A|+|B|=R-1$, and write the product as $\sigma_V^{(A)}\sigma_W^{(B)}$. Combining the standard GKS inequalities with Eq.~\eqref{eq:GKS} gives $ \big\langle \sigma_V^{(A)}\sigma_W^{(B)} \big\rangle_R
\geq
\big\langle \sigma_V^{(A)} \big\rangle_R
\big\langle \sigma_W^{(B)} \big\rangle_R
\geq
\big\langle \sigma_V^{(A)} \big\rangle_{|A|+1}
\big\langle \sigma_W^{(B)} \big\rangle_{|B|+1},$
where replica indices are relabeled as needed in the smaller models. Therefore, if appropriately chosen Ising spin products (and, if necessary, their factors) diagnose the DIPTs for each replica number, Eq.~\eqref{eq:GKS} and this corollary imply the R\'enyi hierarchy: $p_c(R)$ is nondecreasing with integer $R\geq2$.

The key ingredient in our proof of Eq.~\eqref{eq:GKS}, and hence in the resulting R\'enyi hierarchy, is a GKS-type inequality for Ising-type statistical models with nonnegative couplings. For a CSS code under phase-flip decoherence, this condition follows from $J=\frac{1}{2}\log\frac{1-p}{p}\geq0$ for $0<p\leq1/2$, with $p=0$ corresponding to the limit $J\rightarrow+\infty$. This approach naturally suggests extensions to non-CSS codes and more general single-qubit Pauli-type decoherence, whose decoding thresholds and R\'enyi DIPTs also admit descriptions in terms of Ising-type statistical models with random couplings. \cite{dennis2002topological,Flammia2021,FanBaoTopoMemory,BaoFanError_Field_Double,LeeJianXu2023,LeeYouXu2022,SuYangJian2024,Lyons2024StabilizerDecoherence} For R\'enyi index $R\geq2$, the annealed disorder variables can be treated together with the Ising spins as dynamical variables of an enlarged statistical model. If this model has nonnegative couplings and retains the replica structure used above, the same GKS-type argument yields an analogous R\'enyi hierarchy of DIPTs.

\section{$\Zb_2$ topological stabilizer codes with correlated errors}
\label{sec:Z2extended}

In Sec.~\ref{sec:Z2}, we studied the R\'enyi hierarchy in $\Zb_2$ topological stabilizer codes subject to independent decoherence on each qubit. We now extend this discussion to correlated decoherence.

We again consider a $\Zb_2$ topological CSS stabilizer code under phase-flip decoherence. Following Ref.~\onlinecite{Flammia2021}, we consider a correlated decoherence model in which an error chain $\errchn$ occurs with probability
\begin{align}
    P_{\errchn} \propto
    \exp\left\{
        J_0\sum_\mu\varepsilon_\mu
        +J_1\sum_{\langle\mu,\mu'\rangle}
        \varepsilon_\mu\varepsilon_{\mu'}
    \right\},
\end{align}
where $J_0>0$ and $J_1$ controls the correlations between errors. The second sum runs over neighboring pairs of qubits $\mu$ and $\mu'$. Positive $J_1$ favors error bunching, whereas negative $J_1$ favors antibunching.

Following the derivation in Sec.~\ref{sec:Z2}, the replicated Ising-type statistical model associated with the system's $R$th R\'enyi entropy $S_R$ has the partition function
\begin{align}
    \cZ_R \propto
    & \sum_{\{\varepsilon_\mu,\,\sigma_v^{(a)}\}}  \exp\left\{
        J_0 \sum_\mu\varepsilon_\mu
         \left(
            1+\sum_{a=1}^{R-1}
            \prod_{\mu\in v}\sigma_v^{(a)}
        \right) \right.
        \nonumber \\
       & \left. +J_1 \sum_{\langle\mu,\mu'\rangle}
         \varepsilon_\mu \varepsilon_{\mu'}
        \left(
            1+\sum_{a=1}^{R-1}
            \prod_{\mu\in v}\sigma_v^{(a)}  \prod_{\mu'\in v'}\sigma_{v'}^{(a)}
        \right)        
    \right\},
    \label{eq:ZR_extended}
\end{align}
for $R\geq2$.

As $J_0\rightarrow+\infty$ at fixed finite $J_1$, the replicated model favors configurations satisfying $\varepsilon_\mu=1$ and $\prod_{\mu\in v}\sigma_v^{(a)}=1$, including $\varepsilon_\mu=\sigma_v^{(a)}=1$. We refer to the phase connected to this limit as the large-$J_0$ phase. For sufficiently small $J_0$ and $|J_1|$, the model is expected to be in a disordered regime with short-range correlations. The phase diagram of the statistical model $\cZ_R$ has two parameters, and DIPTs, when present, generically occur along one-dimensional curves in the $(J_0,J_1)$ plane. Hence, we reformulate the statement of the R\'enyi hierarchy: at fixed $J_{0,1}$, if $\cZ_R$ lies in the phase continuously connected to its large-$J_0$ limit, then every model with a larger R\'enyi index lies in its corresponding large-$J_0$ phase. In other words, the region occupied by the large-$J_0$ phase in the phase diagram does not shrink as $R$ increases.

For bunching errors, $J_1>0$, all couplings in Eq.~\eqref{eq:ZR_extended} are nonnegative when both $\varepsilon_\mu$ and $\sigma_v^{(a)}$ are treated as dynamical Ising variables when $R\geq 2$. The proof in Sec.~\ref{sec:Z2} therefore applies directly and establishes Eq.~\eqref{eq:GKS} at fixed $J_{0,1}$ for the R\'enyi indices $R>R'\geq2$. Provided that the asymptotic behavior of these correlations characterizes the large-$J_0$ phase and the DIPT out of this phase, Eq.~\eqref{eq:GKS} implies the reformulated R\'enyi hierarchy for R\'enyi index $R\geq2$. Whether the hierarchy extends to the replica limit $R\rightarrow1$ remains an open question for future work.

For $J_1<0$, the replicated model contains negative couplings, so the proof based on GKS-type inequalities no longer applies. This leaves open the question of whether Eq.~\eqref{eq:GKS} and the R\'enyi hierarchy are violated in this regime. However, the failure of the proof does not establish a violation of either statement. We leave this question to future work.

\section{R\'enyi Hierarchy in a decohered rotor model with $\U(1)$ charge conservation}
\label{sec:U1}
In this section, we study the R\'enyi hierarchy of DIPTs in a charge-conserving decohered rotor model relevant to both noisy $\U(1)$-symmetric topological codes and $\U(1)$ strong-to-weak spontaneous symmetry breaking (SWSSB).

Consider a $\U(1)$-symmetric topological code $\cC$ on a 2D square lattice, whose syndromes correspond to the occupation of charged anyonic quasiparticles at each vertex $v$. In the following, we consider a phenomenological model for the charge-conserving decoherence generated by stochastic quasiparticle hoppings. This model has been studied in the contexts of charge-informed quantum error correction \cite{Temkin2025_ChargeInformed} and $\U(1)$ SWSSB \cite{Lessa2025_SWSSB}.

In this phenomenological model, the syndrome is captured by a rotor at each vertex of the square lattice. Its integer-valued number operator $\hat n_v$ measures the quasiparticle occupation, while $e^{\pm\ii\hat\varphi_v}$ raise or lower the occupation by one:
$[\hat n_v,e^{\pm\ii\hat\varphi_v}]=\pm e^{\pm\ii\hat\varphi_v}$. Note that $\hat n_v$ can take negative integer values, representing the occupation by quasiholes. 
Before decoherence, the rotor state is
$\rho_0=\bigotimes_v|0\rangle\langle0|$, representing the vacuum of quasiparticles. 
The decoherence channel is a product of charge-conserving channels on the links of the square lattice:
\begin{align}
    {\cal N}_{\rm tot}&=\prod_\mu{\cal N}_\mu,
    \nonumber\\
   {\rm with}~~  {\cal N}_\mu(\rho)
    &\propto
    \sum_{k_\mu\in\Zb}
    e^{-k_\mu^2/(2\alpha)}
\cdot
    e^{\ii k_\mu\Delta_\mu\hat\varphi}
    \rho
    e^{-\ii k_\mu\Delta_\mu\hat\varphi}.
    \label{eq:AnyonHopping}
\end{align}
Here, $\Delta_\mu$ denotes the oriented lattice derivative along link $\mu$, defined as the variable at its head minus that at its tail. Horizontal links are oriented to the right, and vertical links upward. The operator $e^{\ii k_\mu\Delta_\mu\hat\varphi}$ transfers $k_\mu$ units of charge between its endpoints, with probability proportional to $e^{-k_\mu^2/(2\alpha)}$. The proportionality factor in Eq.~\eqref{eq:AnyonHopping}, which ensures the normalization of probability, is suppressed. The parameter $\alpha>0$ controls the decoherence strength. The limit $\alpha\rightarrow0$ gives the identity channel (without decoherence). This decoherence channel $\cN_{\rm tot}$ respects the strong $\U(1)$ symmetry generated by the total particle number $\sum_v \hat n_v$.

An error chain $E$ is specified by the integers $\{k_\mu\}$, viewed as an integer vector field on the oriented links. The syndrome it generates at vertex $v$ is determined by the lattice divergence $(\Delta k)_v$. Error chains with the same syndrome differ by a divergence-free integer vector field $J$. Their total probability is therefore
\begin{align}
    \cZ_{\{k_\mu\}}
    \propto
    \sum_{J:\,\Delta J=0}
    \exp\left\{
        -\frac{1}{2\alpha}
        \sum_\mu(k_\mu+J_\mu)^2
    \right\}.
    \label{eq:ErrorChainProbU1}
\end{align}
Introducing an angular variable $\theta_v$ at each vertex enforces the constraint $\Delta J=0$ and gives
\begin{align}
    &\cZ_{\{k_\mu\}}
    \nonumber \\
    &\propto
    \int D\theta\sum_J
    \exp\left\{
        -\frac{1}{2\alpha}\sum_\mu(k_\mu+J_\mu)^2
        +\ii\sum_v\theta_v(\Delta J)_v
    \right\}
    \nonumber\\
    &\propto
    \int D\theta\,
    \prod_\mu V_\alpha(\Delta_\mu\theta)\,
    e^{\ii\sum_\mu k_\mu\Delta_\mu\theta},
    \label{eq:SyndromeProbXY}
\end{align}
where $\int D\theta$ denotes $\prod_v\int_0^{2\pi}d\theta_v$.
The function $V_\alpha$ is defined by
\begin{align}
    V_\alpha(\phi)
    &\equiv
    \sum_{m\in\Zb}
    e^{-\frac{\alpha}{2}(\phi-2\pi m)^2}
    \nonumber\\
    &=
    \frac{1}{\sqrt{2\pi\alpha}}
    \sum_{l\in\Zb}
    e^{-l^2/(2\alpha)}e^{-\ii l\phi}.
    \label{eq:V}
\end{align}
The second equality follows from Poisson summation. Note that $V_\alpha$ is an even function and is $2\pi$-periodic. Shifting the summation variable from $J_\mu$ to $J_\mu-k_\mu$ and using Eq.~\eqref{eq:V} gives the last line of Eq.~\eqref{eq:SyndromeProbXY}.

A useful property of this function $V_\alpha$ is
\begin{align}
    V_\alpha(\phi)V_\alpha(\tilde\phi)
    =
    \sum_{s=0,1}
    V_{2\alpha}(\phi_++s\pi)
    V_{2\alpha}(\phi_-+s\pi),
    \label{eq:Vrelation}
\end{align}
where $\phi_\pm=(\phi\pm\tilde\phi)/2$. This identity will be helpful to the discussion of the R\'enyi hierarchy in the DIPTs below.

As in the case of $\Zb_2$ stabilizer codes discussed in Sec.~\ref{sec:Z2}, the DIPT in the rotor system's $R$th R\'enyi entropy is described by a replicated XY-type statistical model with partition function\cite{Temkin2025_ChargeInformed,Lessa2025_SWSSB}
\begin{align}
    \cZ_R & =  \sum_{\{k_\mu\}}  (\cZ_{\{k_\mu\}} )^R  \nonumber \\
    & \propto \sum_{\{k_\mu\}}  \int D\theta  \prod_{a=1}^R \left( \prod_\mu V_\alpha\left(\Delta_\mu\theta^{(a)}\right) ~ e^{\ii \sum_\mu k_\mu \Delta_\mu \theta^{(a)}}\right), \nonumber \\
    & \propto  \int D\theta  \prod_\mu  \left[\prod_{a=1}^{R-1}V_\alpha\left(\Delta_\mu\theta^{(a)}\right)\right] V_\alpha\left(\sum_{a=1}^{R-1} \Delta_\mu\theta^{(a)}\right),
    \label{eq:ZRU1}
\end{align}
where the superscript ${}^{(a)}$ labels replicas, and $\int D\theta$ integrates over all vertices and replicas present in each expression. Technically, $\sum_{\{k_\mu\}}$ should be a sum over one representative $\{k_\mu\}$ of each syndrome class. Formally treating $\sum_{\{k_\mu\}}$ as a summation over all possible $\{k_\mu\}$ configurations does not affect the subsequent analysis. This summation imposes
$\sum_{a=1}^{R}\Delta_\mu\theta^{(a)}=0\ {\rm mod}\ 2\pi$ on every link. Integrating out the $R$th replica gives the last line of Eq. \eqref{eq:ZRU1}, which only contains $R-1$ replicas of angular fields $\theta_v^{(a)}$ and has a $\U(1)^{R-1}$ symmetry. Each $\U(1)$ factor acts as a uniform shift of the $\theta$ field in the corresponding replica.

The relation between $\cZ_R$ and the DIPT in the system's $R$th R\'enyi entropy $S_R$ follows from the orthogonality of distinct syndrome sectors. In the rotor model, each syndrome corresponds to a definite charge distribution and occurs with probability proportional to $\cZ_{\{k_\mu\}}$. Consequently, $S_R=-\frac{1}{R-1}\log\cZ_R$ up to unimportant additive terms. Similar to Sec.~\ref{sec:Z2}, the DIPTs in the decohered rotor model can therefore be studied through the replicated XY-type statistical model $\cZ_R$. The replica limit $R\rightarrow1$ corresponds to the XY-type model with quenched disorder labeled by $\{k_\mu\}$. These disorder variables have already been summed over in Eq.~\eqref{eq:ZRU1}.

In the decoherence-free limit $\alpha\rightarrow0^+$, the function $V_\alpha$ becomes constant, and the angular variables $\theta$ are disordered. Increasing  $\alpha$ favors phase coherence and can drive a Berezinskii--Kosterlitz--Thouless (BKT) transition into a phase with algebraically decaying correlations and the R\'enyi version of the $\U(1)$ quasi-SWSSB\cite{Temkin2025_ChargeInformed,Lessa2025_SWSSB}. We denote the corresponding critical decoherence strength by $\alpha_c(R)$. 

The R\'enyi hierarchy for these DIPTs states that $\alpha_c(R)$ is nondecreasing with $R$. We now establish the R\'enyi hierarchy for $R\geq 2$ by showing that correlations between the $\theta$ variables are nonincreasing with $R$ at fixed $\alpha$. Define
$\Theta=\sum_{v\in V,\,a\in A}q_v^{(a)}\theta_v^{(a)}$,
where $V$ is a set of vertices, $A$ is a set of replicas, and the integer coefficients $q_v^{(a)}$ obey
$\sum_{v\in V}q_v^{(a)}=0$
for each $a\in A$. This neutrality condition makes $\cos\Theta$ invariant under the $\U(1)^{R-1}$ symmetry action. A simple example is
$\cos\Theta=\cos(\theta_v^{(1)}-\theta_{v'}^{(1)})$,
whose expectation value is the two-point correlation function of the angular order parameter $e^{\pm \ii \theta_v^{(a)}}$.

For integer $R\geq2$ and $\Theta$ involving only replicas in $A\subseteq\{1,\ldots,R-1\}$, we will prove the following ordering of correlations:
\begin{align}
    \big\langle\cos\Theta\big\rangle_R
    \geq
    \big\langle\cos\Theta\big\rangle_{R+1}.
    \label{eq:ginibre}
\end{align}
Thus, at fixed $\alpha$, these correlations are stronger for smaller $R$. Since their long-distance behavior diagnoses the BKT transition, the replicated statistical model $\cZ_R$ with smaller $R$ leaves the disordered phase at an equal or smaller value of $\alpha_c$. This implies that $\alpha_c(R)$ is nondecreasing with $R$.

To prove Eq.~\eqref{eq:ginibre}, we deform $\cZ_{R+1}$ by introducing a field $h\geq0$ that biases the angular field $\theta_v^{(R)}$ in the replicated statistical model with R\'enyi index $R+1$:
\begin{align}
    &\cZ_{R+1,h}   
    \propto  \int \! D\theta  \prod_\mu  \left[\prod_{a=1}^{R}V_\alpha\left(\Delta_\mu\theta^{(a)}\right)\right] V_\alpha\left(\sum_{a=1}^{R} \Delta_\mu\theta^{(a)}\right) 
    \nonumber \\
    & ~~~~~~~~~~~~~~~~~~\times \exp\left( \sum_v h\cos \theta_v^{(R)} \right).
\end{align}
At $h=0$, the deformed model recovers the original statistical model with R\'enyi index $R+1$. As $h\rightarrow+\infty$, the angles $\theta_v^{(R)}$ are pinned to zero modulo $2\pi$, and $\cZ_{R+1,h}$ reduces to the statistical model $\cZ_R$ with  R\'enyi index $R$ (up to an overall normalization factor). We denote expectation values in the deformed model by $\langle \,\cdot \,\rangle_h$, suppressing the R\'enyi index.

For Eq. \eqref{eq:ginibre}, it suffices to prove
\begin{align}
    &\frac{d}{dh}
    \big\langle\cos\Theta\big\rangle_h
    \nonumber\\
    &=
    \sum_u
    \Big[
        \big\langle
            \cos\Theta\cos\theta_u^{(R)}
        \big\rangle_h
        -
        \big\langle\cos\Theta\big\rangle_h
        \big\langle\cos\theta_u^{(R)}\big\rangle_h
    \Big]
    \geq0.
    \label{eq:FlucDsp}
\end{align}
In fact, we will show that each connected correlation in this sum is nonnegative. Following the strategy of Ginibre-type inequalities \cite{ginibre1970general} (which generalizes the GKS inequalities), we write
\begin{widetext}
\begin{align}
&\big\langle\cos \Theta  \cos \theta_u^{(R)} \big\rangle_{h} - \big\langle \cos \Theta \big\rangle_{h} \big\langle\cos \theta_u^{(R)} \big\rangle_{h} 
\nonumber \\
&= \frac{1}{2\cZ_{R+1,h}^2} \int  D\theta \int  D\tilde\theta 
\left(\cos \Theta - \cos \tilde\Theta \right)
\left(\cos \theta_u^{(R)} - \cos \tilde\theta_u^{(R)} \right)
\nonumber \\
& ~~~~~~~~~~~~~~ 
\times \prod_\mu  \left[\prod_{a=1}^{R}V_\alpha\left(\Delta_\mu\theta^{(a)}\right)\right]
V_\alpha\left(\sum_{a=1}^{R} \Delta_\mu\theta^{(a)}\right)  
\exp\left( \sum_v h\cos \theta_v^{(R)} \right)
\nonumber \\
& ~~~~~~~~~~~~~~ 
\times \prod_\mu  \left[\prod_{a=1}^{R}V_\alpha\left(\Delta_\mu \tilde\theta^{(a)}\right)\right] 
V_\alpha\left(\sum_{a=1}^{R} \Delta_\mu\tilde\theta^{(a)}\right)  
\exp\left( \sum_v h\cos \tilde\theta_v^{(R)} \right)
\nonumber \\
& = \frac{1}{2\cZ_{R+1,h}^2} \int  D\theta_+ \int  D\theta_- 
\left(2\sin \theta_{u,+}^{(R)}  \sin \theta_{u,-}^{(R)}  \right)
\left(2\sin \Theta_{+}  \sin \Theta_{-}  \right)
\nonumber \\
& ~~~~ \times \sum_{\{s_\mu^{(a)} = 0,1\}} 
\left\{ \prod_\mu  \left[\prod_{a=1}^{R}V_{2\alpha}\left(\Delta_\mu\theta^{(a)}_+ + s_\mu^{(a)} \pi \right)\right] 
V_{2\alpha}\left(\sum_{a=1}^{R} \Delta_\mu\theta^{(a)}_+ + s_\mu^{(R+1)} \pi\right)   \right.
\nonumber \\
& ~~~~~~~~~~~~~\left. \times \prod_\mu  \left[\prod_{a=1}^{R}V_{2\alpha}\left(\Delta_\mu \theta^{(a)}_-  + s_\mu^{(a)} \pi \right)\right] 
V_{2\alpha}\left(\sum_{a=1}^{R} \Delta_\mu \theta^{(a)}_- + s_\mu^{(R+1)} \pi \right)  
\exp\left( 2\sum_v h\cos \theta_{v,+}^{(R)}  \cos \theta_{v,-}^{(R)} \right) \right\} \geq 0.
\label{eq:CovProof}
\end{align}
\end{widetext}
The first equality is a direct rewriting of the connected correlation in the first line. For the second, we introduce
$\theta_{v,\pm}^{(a)}=(\theta_v^{(a)}\pm\tilde\theta_v^{(a)})/2$
and
$\Theta_{\pm}=\sum_{v\in V,\,a\in A}q_v^{(a)}\theta_{v,\pm}^{(a)}$,
and apply Eq.~\eqref{eq:Vrelation} to each pair of $V_{\alpha}$ factors. For every link, the binary variables $s_\mu^{(a)}$ with $a=1,\ldots, R+1$ are independent. 
For each pair of variables $\theta$ and $\tilde\theta$, we have also replaced $\int_0^{2\pi}d\theta\int_0^{2\pi}d\tilde\theta$ by $\int_0^{2\pi}d\theta_+\int_0^{2\pi}d\theta_-$. This replacement is valid because the integrand is $2\pi$-periodic in both $\theta$ and $\tilde\theta$. Under $\theta=\theta_++\theta_-$ and $\tilde\theta=\theta_+-\theta_-$, the integration domain $(\theta_+,\theta_-)\in[0,2\pi)^2$ covers the original integration domain $(\theta ,\tilde \theta)\in[0,2\pi)^2$ twice, while the Jacobian has magnitude $2$. These factors cancel. With these operations and standard trigonometric identities, we obtain the second expression in Eq.~\eqref{eq:CovProof}.

To establish the nonnegativity of Eq.~\eqref{eq:CovProof}, we expand 
\begin{align}
    &\exp\left(
        2h\sum_v
        \cos\theta_{v,+}^{(R)}
        \cos\theta_{v,-}^{(R)}
    \right)
    \nonumber \\
    &~~~~~~~=
    \prod_v\sum_{m_v=0}^{\infty}
    \frac{(2h)^{m_v}}{m_v!}
    \left(
        \cos\theta_{v,+}^{(R)}
        \cos\theta_{v,-}^{(R)}
    \right)^{m_v}.
\end{align}
Every coefficient is nonnegative for $h\geq0$. For each fixed set of binary variables $\{s_\mu^{(a)}\}$ and powers $\{m_v\}$, the $\theta_+$ and $\theta_-$ integrals factorize into identical real integrals. Their product is therefore a square, proving Eq.~\eqref{eq:CovProof} and hence Eq.~\eqref{eq:FlucDsp}.

Taking $h$ from zero to $+\infty$ proves Eq.~\eqref{eq:ginibre} for the correlation functions. Since the asymptotic behavior of these correlation functions characterizes the DIPTs of the BKT type, we obtain $\alpha_c(R)\leq\alpha_c(R+1)$ for $R\geq2$. This establishes the R\'enyi hierarchy of DIPTs in the decohered rotor model that describes the effect of the decoherence channel in Eq.~\eqref{eq:AnyonHopping} acting on the quasiparticle vacuum $\rho_0=\bigotimes_v|0\rangle\langle0|$. The analysis above can be straightforwardly generalized to higher dimension decohered rotor models where DIPTs into $\U(1)$ SWSSB phases with true long-range order can emerge.  

An interesting direction for future work is to investigate whether this R\'enyi hierarchy extends to the replica limit $R\rightarrow1$, both in the present model and in more general $\U(1)$-symmetric decohered systems.

\section{Conclusion and Outlook}
\label{sec:conclusion}

In this paper, we identify a general mechanism for the R\'enyi hierarchy of DIPTs in decohered $\Zb_2$ topological stabilizer codes and a $\U(1)$-symmetric decohered rotor model. The common ingredient is an ordering of correlations in the replicated random statistical models that describe the system's R\'enyi entropies $S_R$ and their associated DIPTs. 

In Sec.~\ref{sec:Z2}, for CSS topological codes under independent phase-flip decoherence, we establish the correlation ordering Eq.~\eqref{eq:GKS} using GKS-type inequalities for classical Ising spin models with nonnegative couplings. This ordering states that, at fixed decoherence strength, the expectation values of any fixed product of Ising spins in the replicated statistical models are nondecreasing with integer R\'enyi index $R\geq2$. Assuming that the long-distance or large-loop behavior of suitably chosen spin correlations diagnoses the DIPTs, this ordering implies that the critical decoherence strength $p_c(R)$ is also nondecreasing with $R$. Through the correspondence between R\'enyi-$R$ DIPTs and quantum error correction with postselection across $R$ copies of the decohered code~\cite{Mong2025_ReplicaTopo}, this result also implies a hierarchy of decoding thresholds in the relevant codes. These results suggest extensions to non-CSS codes and more general single-qubit Pauli noise, provided the corresponding replicated statistical models retain nonnegative couplings and a similar replica structure.

In Sec.~\ref{sec:Z2extended}, we extend the analysis to correlated phase-flip errors, for which Eq.~\eqref{eq:GKS} remains valid in the regime with bunching errors ($J_1>0$). Within this regime, the region of the large-$J_0$ phase in the $(J_0,J_1)$ phase diagram cannot shrink as the R\'enyi index $R\geq2$ increases, provided the correlations characterize that phase. Antibunching errors ($J_1<0$) introduce negative couplings to the replicated statistical models that invalidate the GKS-based proof.

In Sec.~\ref{sec:U1}, we study a decohered rotor model with strong $\U(1)$ symmetry, previously considered in the contexts of charge-informed quantum error correction and $\U(1)$ SWSSB. We apply a Ginibre-type generalization of the GKS argument to establish the correlation ordering in Eq.~\eqref{eq:ginibre} for the replicated XY-type statistical models that describe the rotor model's DIPTs. At fixed decoherence strength $\alpha$, the angular correlations are nonincreasing with the R\'enyi index $R\geq2$. Since their asymptotic behavior characterizes the DIPTs, this ordering implies that the critical decoherence strength $\alpha_c(R)$ for the R\'enyi-$R$ DIPT is nondecreasing with $R$.

Several directions remain open. First, Eqs.~\eqref{eq:GKS} and \eqref{eq:ginibre} are established only for integer R\'enyi indices $R\geq2$. An important question is whether these inequalities and the DIPT R\'enyi hierarchy extend to the replica limit $R\rightarrow1$. In the underlying random statistical models, this limit is associated with quenched disorder, whereas the disorder is effectively annealed for $R\geq2$. For the $\Zb_2$ stabilizer codes, establishing inequalities between correlations with quenched disorder and their finite-replica counterparts would extend our arguments of the R\'enyi hierarchy of DIPTs to the von Neumann limit. For the decohered rotor model, Ref.~\onlinecite{Temkin2025_ChargeInformed} showed a peculiar feature that angular correlations such as $\langle \cos(\theta_v^{(1)}-\theta_{v'}^{(1)}) \rangle_{R\rightarrow1}$ are identically equal to one and therefore cannot diagnose the DIPT. Extending the argument for the R\'enyi hierarchy of DIPTs to the replica limit $R\rightarrow1$ thus requires a different approach.

Second, the antibunching regime of Sec.~\ref{sec:Z2extended} warrants further investigation. The failure of the GKS-based proof does not establish a violation of either Eq.~\eqref{eq:GKS} or the R\'enyi hierarchy. Determining whether the correlation ordering or the R\'enyi hierarchy of phase boundaries actually fails in this regime would clarify the scope of the proposed mechanism.

Finally, it would be interesting to extend the Ginibre-based argument beyond the rotor model to other systems with decoherence that conserve $\U(1)$ charge. Examples include the decohered spin-$1/2$ systems studied in Ref.~\onlinecite{Hauser2026SWSSBHydro} and the fermionic systems studied in Refs.~\onlinecite{XuBloch2026ExpSWSSB,WangEhud2025NoisyFQH,SuSarmaXu2025SL}. Investigating whether analogous correlation inequalities and R\'enyi hierarchies hold in these systems could deepen our understanding of R\'enyi-$R$ $\U(1)$ SWSSB in broader settings.

\section*{Acknowledgement}
We thank Jong Yeon Lee, Kaixiang Su, and Cenke Xu for helpful discussions. 
Z.Y. and C.-M.J. are supported in part by the Alfred P. Sloan Foundation through a Sloan Research Fellowship. C.-M.J. is supported in part by the National Science Foundation under Award No. PHY-2609907. We acknowledge the use of ChatGPT (OpenAI) for language polishing and proofreading of this manuscript. The authors reviewed and edited all generated text and take full responsibility for the final content.

~

{\it Note added} - While preparing this manuscript, we became aware of an independent related work~\cite{VijayLee2026Hierarchy}, which also investigates the R\'enyi hierarchy of DIPTs in $\Zb_2$ stabilizer codes under Pauli noise. Ref.~\onlinecite{VijayLee2026Hierarchy} establishes an ordering of the von Neumann and R\'enyi coherent information for codes under Pauli noise from independent Bernoulli events. Its proof also relies on GKS-type inequalities for R\'enyi indices $R\geq2$. When these coherent information measures diagnose the codes' DIPTs, this ordering implies the R\'enyi hierarchy. Here, we address the hierarchy for $R\geq2$ directly through the ordering of correlation functions in replicated statistical models. Our analysis includes both $\Zb_2$ stabilizer codes and a decohered rotor model with strong $\U(1)$ symmetry.

\bibliography{ref}

@ARTICLE{Lee2026Scrambling,
       author = {{Lee}, Jong Yeon},
        title = "{Charge Scrambling in Strong-to-Weak Spontaneous Symmetry Breaking}",
      journal = {arXiv e-prints},
         year = 2026,
        month = may,
          eid = {arXiv:2605.05288},
        pages = {arXiv:2605.05288},
          doi = {10.48550/arXiv.2605.05288},
archivePrefix = {arXiv},
       eprint = {2605.05288},
 primaryClass = {cond-mat.stat-mech},
       adsurl = {https://ui.adsabs.harvard.edu/abs/2026arXiv260505288L}
}

@ARTICLE{OgunnaikeLee2023Lind,
       author = {{Ogunnaike}, Olumakinde and {Feldmeier}, Johannes and {Lee}, Jong Yeon},
        title = "{Unifying Emergent Hydrodynamics and Lindbladian Low-Energy Spectra across Symmetries, Constraints, and Long-Range Interactions}",
      journal = {\prl},
         year = 2023,
        month = dec,
       volume = {131},
       number = {22},
          eid = {220403},
        pages = {220403},
          doi = {10.1103/PhysRevLett.131.220403},
archivePrefix = {arXiv},
       eprint = {2304.13028},
 primaryClass = {cond-mat.str-el},
       adsurl = {https://ui.adsabs.harvard.edu/abs/2023PhRvL.131v0403O}
}

@article{Kubica2018ColorCodeThresholds,
  author = {Kubica, Aleksander and Beverland, Michael E. and
            Brand{\~a}o, Fernando and Preskill, John and Svore, Krysta M.},
  title = {Three-Dimensional Color Code Thresholds via Statistical-Mechanical Mapping},
  journal = {Physical Review Letters},
  volume = {120},
  number = {18},
  pages = {180501},
  year = {2018},
  doi = {10.1103/PhysRevLett.120.180501},
  eprint = {1708.07131},
  archivePrefix = {arXiv}
}

@article{KimAltmanLee2026SYKThreshold,
  title = {Error threshold of {Sachdev-Ye-Kitaev} models from
           strong-to-weak parity symmetry breaking},
  author = {Kim, Jaewon and Altman, Ehud and Lee, Jong Yeon},
  journal = {Physical Review B},
  volume = {114},
  number = {5},
  pages = {L051101},
  year = {2026},
  month = {July},
  publisher = {American Physical Society},
  doi = {10.1103/srf2-1f6d},
  url = {https://doi.org/10.1103/srf2-1f6d},
  eprint = {2410.24225},
  archivePrefix = {arXiv},
  primaryClass = {quant-ph}
}

@article{BehrendsBeri2025BeyondPauli,
  title = {The Surface Code beyond Pauli Channels: Logical Noise Coherence, Information-Theoretic Measures, and Errorfield-Double Phenomenology},
  author = {Behrends, Jan and B{\'e}ri, Benjamin},
  journal = {PRX Quantum},
  volume = {6},
  issue = {4},
  pages = {040350},
  numpages = {22},
  year = {2025},
  month = {Dec},
  publisher = {American Physical Society},
  doi = {10.1103/psf5-b6j2},
  url = {https://doi.org/10.1103/psf5-b6j2},
  eprint = {2412.21055},
  archivePrefix = {arXiv},
  primaryClass = {quant-ph}
}

@article{Lyons2024StabilizerDecoherence,
  title = {Understanding Stabilizer Codes Under Local Decoherence Through a General Statistical Mechanics Mapping},
  author = {Lyons, Anasuya},
  journal = {arXiv e-prints},
  year = {2024},
  month = {Mar},
  pages = {arXiv:2403.03955},
  doi = {10.48550/arXiv.2403.03955},
  url = {https://arxiv.org/abs/2403.03955},
  eprint = {2403.03955},
  archivePrefix = {arXiv},
  primaryClass = {quant-ph}
}

@article{Colmenarez2024CoherentThresholds,
  title = {Accurate optimal quantum error correction thresholds from coherent information},
  author = {Colmenarez, Luis and Huang, Ze-Min and Diehl, Sebastian and M{\"u}ller, Markus},
  journal = {Phys. Rev. Res.},
  volume = {6},
  issue = {4},
  pages = {L042014},
  numpages = {6},
  year = {2024},
  month = {Oct},
  publisher = {American Physical Society},
  doi = {10.1103/PhysRevResearch.6.L042014},
  url = {https://doi.org/10.1103/PhysRevResearch.6.L042014},
  eprint = {2312.06664},
  archivePrefix = {arXiv},
  primaryClass = {quant-ph}
}

@article{NiwaLee2025CSSCoherentInformation,
  title = {Coherent information for Calderbank-Shor-Steane codes under decoherence},
  author = {Niwa, Ryotaro and Lee, Jong Yeon},
  journal = {Phys. Rev. A},
  volume = {111},
  issue = {3},
  pages = {032402},
  numpages = {13},
  year = {2025},
  month = {Mar},
  publisher = {American Physical Society},
  doi = {10.1103/PhysRevA.111.032402},
  url = {https://doi.org/10.1103/PhysRevA.111.032402},
  eprint = {2407.02564},
  archivePrefix = {arXiv},
  primaryClass = {quant-ph}
}

@article{Lee2025ExactCoherentInformation,
  title = {Exact Calculations of Coherent Information for Toric Codes under Decoherence: Identifying the Fundamental Error Threshold},
  author = {Lee, Jong Yeon},
  journal = {Phys. Rev. Lett.},
  volume = {134},
  issue = {25},
  pages = {250601},
  numpages = {7},
  year = {2025},
  month = {Jun},
  publisher = {American Physical Society},
  doi = {10.1103/hlfh-86yz},
  url = {https://doi.org/10.1103/hlfh-86yz},
  eprint = {2402.16937},
  archivePrefix = {arXiv},
  primaryClass = {cond-mat.stat-mech}
}

@article{Wang2025SelfDualCriticality,
  title = {Decoherence-induced self-dual criticality in topological states of matter},
  author = {Wang, Qingyuan and Vasseur, Romain and Trebst, Simon and Ludwig, Andreas W. W. and Zhu, Guo-Yi},
  journal = {arXiv e-prints},
  year = {2025},
  month = {Feb},
  pages = {arXiv:2502.14034},
  doi = {10.48550/arXiv.2502.14034},
  url = {https://arxiv.org/abs/2502.14034},
  eprint = {2502.14034},
  archivePrefix = {arXiv},
  primaryClass = {quant-ph}
}

@article{KellySherman1968,
    author = {Kelly, D. G. and Sherman, S.},
    title = {General Griffiths' Inequalities on Correlations in Ising Ferromagnets},
    journal = {Journal of Mathematical Physics},
    volume = {9},
    number = {3},
    pages = {466-484},
    year = {1968},
    month = {03},
    issn = {0022-2488},
    doi = {10.1063/1.1664600},
    url = {https://doi.org/10.1063/1.1664600},
}

@article{Griffiths1967,
    author = {Griffiths, Robert B.},
    title = {Correlations in Ising Ferromagnets. I},
    journal = {Journal of Mathematical Physics},
    volume = {8},
    number = {3},
    pages = {478-483},
    year = {1967},
    month = {03},
    issn = {0022-2488},
    doi = {10.1063/1.1705219},
    url = {https://doi.org/10.1063/1.1705219},
}

@article{ginibre1970general,
  title={General formulation of Griffiths' inequalities},
  author={Ginibre, Jean},
  journal={Communications in mathematical physics},
  volume={16},
  number={4},
  pages={310--328},
  year={1970},
  publisher={Springer}
}

@article{Lessa2025_SWSSB,
  title = {Strong-to-Weak Spontaneous Symmetry Breaking in Mixed Quantum States},
  author = {Lessa, Leonardo A. and Ma, Ruochen and Zhang, Jian-Hao and Bi, Zhen and Cheng, Meng and Wang, Chong},
  journal = {PRX Quantum},
  volume = {6},
  issue = {1},
  pages = {010344},
  numpages = {24},
  year = {2025},
  month = {Mar},
  publisher = {American Physical Society},
  doi = {10.1103/PRXQuantum.6.010344},
  url = {https://link.aps.org/doi/10.1103/PRXQuantum.6.010344}
}

@ARTICLE{Mong2025_ReplicaTopo,
       author = {{Li}, Zhuan and {Mong}, Roger S.~K.},
        title = "{Replica topological order in quantum mixed states and quantum error correction}",
      journal = {\prb},
         year = 2025,
        month = mar,
       volume = {111},
       number = {12},
          eid = {125106},
        pages = {125106},
          doi = {10.1103/PhysRevB.111.125106},
archivePrefix = {arXiv},
       eprint = {2402.09516},
 primaryClass = {quant-ph},
       adsurl = {https://ui.adsabs.harvard.edu/abs/2025PhRvB.111l5106L}
}

@ARTICLE{Temkin2025_ChargeInformed,
       author = {{Temkin}, Vlad and {Weinstein}, Zack and {Fan}, Ruihua and {Podolsky}, Daniel and {Altman}, Ehud},
        title = "{Charge-Informed Quantum Error Correction}",
      journal = {arXiv e-prints},
         year = 2025,
        month = dec,
          eid = {arXiv:2512.22119},
        pages = {arXiv:2512.22119},
          doi = {10.48550/arXiv.2512.22119},
archivePrefix = {arXiv},
       eprint = {2512.22119},
 primaryClass = {quant-ph},
       adsurl = {https://ui.adsabs.harvard.edu/abs/2025arXiv251222119T}
}

@ARTICLE{SuYangJian2024,
       author = {{Su}, Kaixiang and {Yang}, Zhou and {Jian}, Chao-Ming},
        title = "{Tapestry of dualities in decohered quantum error correction codes}",
      journal = {\prb},
         year = 2024,
        month = aug,
       volume = {110},
       number = {8},
          eid = {085158},
        pages = {085158},
          doi = {10.1103/PhysRevB.110.085158},
archivePrefix = {arXiv},
       eprint = {2401.17359},
 primaryClass = {cond-mat.str-el},
       adsurl = {https://ui.adsabs.harvard.edu/abs/2024PhRvB.110h5158S}
}

@ARTICLE{Grover2024Separability,
       author = {{Chen}, Yu-Hsueh and {Grover}, Tarun},
        title = "{Separability Transitions in Topological States Induced by Local Decoherence}",
      journal = {\prl},
         year = 2024,
        month = apr,
       volume = {132},
       number = {17},
          eid = {170602},
        pages = {170602},
          doi = {10.1103/PhysRevLett.132.170602},
archivePrefix = {arXiv},
       eprint = {2309.11879},
 primaryClass = {quant-ph},
       adsurl = {https://ui.adsabs.harvard.edu/abs/2024PhRvL.132q0602C}
}

@article{canossa2023exotic,
   title={Exotic symmetry breaking properties of self-dual fracton spin models},
   volume={6},
   ISSN={2643-1564},
   url={http://dx.doi.org/10.1103/PhysRevResearch.6.013304},
   DOI={10.1103/physrevresearch.6.013304},
   number={1},
   journal={Physical Review Research},
   publisher={American Physical Society (APS)},
   author={Canossa, Giovanni and Pollet, Lode and Martin-Delgado, Miguel A. and Song, Hao and Liu, Ke},
   year={2024},
   month=Mar }

@ARTICLE{LeeYouXu2022,
       author = {{Lee}, Jong Yeon and {You}, Yi-Zhuang and {Xu}, Cenke},
        title = "{Symmetry protected topological phases under decoherence}",
      journal = {arXiv e-prints},
         year = 2022,
        month = oct,
          eid = {arXiv:2210.16323},
        pages = {arXiv:2210.16323},
          doi = {10.48550/arXiv.2210.16323},
archivePrefix = {arXiv},
       eprint = {2210.16323},
 primaryClass = {cond-mat.str-el},
       adsurl = {https://ui.adsabs.harvard.edu/abs/2022arXiv221016323L}
}

@article{Flammia2021,
	doi = {10.4171/aihpd/105},
  
	url = {https://doi.org/10.4171%2Faihpd%2F105},
  
	year = 2021,
	month = {may},
  
	publisher = {European Mathematical Society - {EMS} - Publishing House {GmbH}
},
  
	volume = {8},
  
	number = {2},
  
	pages = {269--321},
  
	author = {Christopher T. Chubb and Steven T. Flammia},
  
	title = {Statistical mechanical models for quantum codes with correlated noise},
  
	journal = {Annales de l'Institut Henri Poincar{\'{e}} D}
}

@article{dennis2002topological,
  title={Topological quantum memory},
  author={Dennis, Eric and Kitaev, Alexei and Landahl, Andrew and Preskill, John},
  journal={Journal of Mathematical Physics},
  volume={43},
  number={9},
  pages={4452--4505},
  year={2002},
  publisher={American Institute of Physics}
}

@article{Matin-DelgadoColorCodes,
  title = {Error Threshold for Color Codes and Random Three-Body Ising Models},
  author = {Katzgraber, Helmut G. and Bombin, H. and Martin-Delgado, M. A.},
  journal = {Phys. Rev. Lett.},
  volume = {103},
  issue = {9},
  pages = {090501},
  numpages = {4},
  year = {2009},
  month = {Aug},
  publisher = {American Physical Society},
  doi = {10.1103/PhysRevLett.103.090501},
  url = {https://link.aps.org/doi/10.1103/PhysRevLett.103.090501}
}

@article{Preskill_3dToricCodeThreshold,
	doi = {10.1016/s0003-4916(02)00019-2},
  
	url = {https://doi.org/10.1016%2Fs0003-4916%2802%2900019-2},
  
	year = 2003,
	month = {jan},
  
	publisher = {Elsevier {BV}
},
  
	volume = {303},
  
	number = {1},
  
	pages = {31--58},
  
	author = {Chenyang Wang and Jim Harrington and John Preskill},
  
	title = {Confinement-Higgs transition in a disordered gauge theory and the accuracy threshold for quantum memory},
  
	journal = {Annals of Physics}
}

@article{Martin-Delgado_FractonThreshold,
  title = {Optimal Thresholds for Fracton Codes and Random Spin Models with Subsystem Symmetry},
  author = {Song, Hao and Sch\"onmeier-Kromer, Janik and Liu, Ke and Viyuela, Oscar and Pollet, Lode and Martin-Delgado, M. A.},
  journal = {Phys. Rev. Lett.},
  volume = {129},
  issue = {23},
  pages = {230502},
  numpages = {7},
  year = {2022},
  month = {Nov},
  publisher = {American Physical Society},
  doi = {10.1103/PhysRevLett.129.230502},
  url = {https://link.aps.org/doi/10.1103/PhysRevLett.129.230502}
}

@ARTICLE{FanBaoTopoMemory,
       author = {{Fan}, Ruihua and {Bao}, Yimu and {Altman}, Ehud and {Vishwanath}, Ashvin},
        title = "{Diagnostics of Mixed-State Topological Order and Breakdown of Quantum Memory}",
      journal = {PRX Quantum},
         year = 2024,
        month = may,
       volume = {5},
       number = {2},
          eid = {020343},
        pages = {020343},
          doi = {10.1103/PRXQuantum.5.020343},
archivePrefix = {arXiv},
       eprint = {2301.05689},
 primaryClass = {quant-ph},
       adsurl = {https://ui.adsabs.harvard.edu/abs/2024PRXQ....5b0343F}
}

@ARTICLE{Hauser2026SWSSBHydro,
       author = {{Hauser}, Jacob and {Su}, Kaixiang and {Ha}, Hyunsoo and {Lloyd}, Jerome and {Kiely}, Thomas G. and {Vasseur}, Romain and {Gopalakrishnan}, Sarang and {Xu}, Cenke and {Fisher}, Matthew P.~A.},
        title = "{Strong-to-Weak Symmetry Breaking in Open Quantum Systems: From Discrete Particles to Continuum Hydrodynamics}",
      journal = {arXiv e-prints},
         year = 2026,
        month = feb,
          eid = {arXiv:2602.16045},
        pages = {arXiv:2602.16045},
          doi = {10.48550/arXiv.2602.16045},
archivePrefix = {arXiv},
       eprint = {2602.16045},
 primaryClass = {quant-ph},
       adsurl = {https://ui.adsabs.harvard.edu/abs/2026arXiv260216045H}
}

@ARTICLE{XuBloch2026ExpSWSSB,
       author = {{Wang}, Si and {Kiely}, Thomas G. and {Tell}, Dorothee and {Obermeyer}, Johannes and {Barendregt}, Marnix and {Bojovi{\'c}}, Petar and {Preiss}, Philipp M. and {Sarma}, Abhijat and {Franz}, Titus and {Fisher}, Matthew P.~A. and {Xu}, Cenke and {Bloch}, Immanuel},
        title = "{Observation of Strong-to-Weak Spontaneous Symmetry Breaking in a Dephased Fermi Gas}",
      journal = {arXiv e-prints},
         year = 2026,
        month = apr,
          eid = {arXiv:2604.16137},
        pages = {arXiv:2604.16137},
          doi = {10.48550/arXiv.2604.16137},
archivePrefix = {arXiv},
       eprint = {2604.16137},
 primaryClass = {cond-mat.quant-gas},
       adsurl = {https://ui.adsabs.harvard.edu/abs/2026arXiv260416137W}
}

@ARTICLE{You2024PurificationSWSSB,
       author = {{Sala}, Pablo and {Gopalakrishnan}, Sarang and {Oshikawa}, Masaki and {You}, Yizhi},
        title = "{Spontaneous strong symmetry breaking in open systems: Purification perspective}",
      journal = {\prb},
         year = 2024,
        month = oct,
       volume = {110},
       number = {15},
          eid = {155150},
        pages = {155150},
          doi = {10.1103/PhysRevB.110.155150},
archivePrefix = {arXiv},
       eprint = {2405.02402},
 primaryClass = {quant-ph},
       adsurl = {https://ui.adsabs.harvard.edu/abs/2024PhRvB.110o5150S}
}

@article{SuSarmaXu2025SL,
  title = {Spin Liquid and Superconductivity Emerging from Steady States and Measurements},
  author = {Su, Kaixiang and Sarma, Abhijat and Bintz, Marcus and Kiely, Thomas and Bao, Yimu and Fisher, Matthew P. A. and Xu, Cenke},
  journal = {Phys. Rev. Lett.},
  volume = {135},
  issue = {5},
  pages = {050403},
  numpages = {6},
  year = {2025},
  month = {Aug},
  publisher = {American Physical Society},
  doi = {10.1103/pjs3-14cc},
  url = {https://link.aps.org/doi/10.1103/pjs3-14cc}
}

@ARTICLE{VijayLee2026Hierarchy,
       author = {{Vijay}, Akash and {Colmenarez}, Luis and {Lee}, Jong Yeon},
        title = "{Hierarchy of R{\'e}nyi Coherent Information in Stabilizer Codes}",
      journal = {arXiv e-prints},
         year = 2026,
        month = sep,
          eid = {arXiv:2609.11930},
        pages = {arXiv:2609.11930},
          doi = {10.48550/arXiv.2609.11930},
archivePrefix = {arXiv},
       eprint = {2609.11930},
 primaryClass = {quant-ph},
       adsurl = {https://ui.adsabs.harvard.edu/abs/2026arXiv260911930V}
}

@ARTICLE{WangEhud2025NoisyFQH,
       author = {{Wang}, Zijian and {Fan}, Ruihua and {Wang}, Tianle and {Garratt}, Samuel J. and {Altman}, Ehud},
        title = "{Fractional quantum Hall states under density decoherence}",
      journal = {arXiv e-prints},
         year = 2025,
        month = oct,
          eid = {arXiv:2510.08490},
        pages = {arXiv:2510.08490},
          doi = {10.48550/arXiv.2510.08490},
archivePrefix = {arXiv},
       eprint = {2510.08490},
 primaryClass = {cond-mat.str-el},
       adsurl = {https://ui.adsabs.harvard.edu/abs/2025arXiv251008490W}
}

@article{SangZouHsieh2024RG,
  title = {Mixed-State Quantum Phases: Renormalization and Quantum Error Correction},
  author = {Sang, Shengqi and Zou, Yijian and Hsieh, Timothy H.},
  journal = {Phys. Rev. X},
  volume = {14},
  issue = {3},
  pages = {031044},
  numpages = {24},
  year = {2024},
  month = {Sep},
  publisher = {American Physical Society},
  doi = {10.1103/PhysRevX.14.031044},
  url = {https://link.aps.org/doi/10.1103/PhysRevX.14.031044}
}

@article{MaWang2025ASPT,
  title = {Topological Phases with Average Symmetries: The Decohered, the Disordered, and the Intrinsic},
  author = {Ma, Ruochen and Zhang, Jian-Hao and Bi, Zhen and Cheng, Meng and Wang, Chong},
  journal = {Phys. Rev. X},
  volume = {15},
  issue = {2},
  pages = {021062},
  numpages = {33},
  year = {2025},
  month = {May},
  publisher = {American Physical Society},
  doi = {10.1103/PhysRevX.15.021062},
  url = {https://link.aps.org/doi/10.1103/PhysRevX.15.021062}
}

@ARTICLE{HuangLucas2025,
       author = {{Huang}, Xiaoyang and {Qi}, Marvin and {Zhang}, Jian-Hao and {Lucas}, Andrew},
        title = "{Hydrodynamics as the effective field theory of strong-to-weak spontaneous symmetry breaking}",
      journal = {\prb},
         year = 2025,
        month = mar,
       volume = {111},
       number = {12},
          eid = {125147},
        pages = {125147},
          doi = {10.1103/PhysRevB.111.125147},
archivePrefix = {arXiv},
       eprint = {2407.08760},
 primaryClass = {cond-mat.str-el},
       adsurl = {https://ui.adsabs.harvard.edu/abs/2025PhRvB.111l5147H}
}

@article{Zhang2025HigherFormSWSSB,
  title = {Strong-to-weak spontaneous breaking of 1-form symmetry and intrinsically mixed topological order},
  author = {Zhang, Carolyn and Xu, Yichen and Zhang, Jian-Hao and Xu, Cenke and Bi, Zhen and Luo, Zhu-Xi},
  journal = {Phys. Rev. B},
  volume = {111},
  issue = {11},
  pages = {115137},
  numpages = {28},
  year = {2025},
  month = {Mar},
  publisher = {American Physical Society},
  doi = {10.1103/PhysRevB.111.115137},
  url = {https://link.aps.org/doi/10.1103/PhysRevB.111.115137}
}

@ARTICLE{BaoFanError_Field_Double,
       author = {{Bao}, Yimu and {Fan}, Ruihua and {Vishwanath}, Ashvin and {Altman}, Ehud},
        title = "{Mixed-State Topological Order and the Errorfield Double Formulation of Decoherence-Induced Transitions}",
      journal = {\prl},
         year = 2026,
        month = jun,
       volume = {136},
       number = {22},
          eid = {220402},
        pages = {220402},
          doi = {10.1103/6f98-tvb8},
archivePrefix = {arXiv},
       eprint = {2301.05687},
 primaryClass = {quant-ph},
       adsurl = {https://ui.adsabs.harvard.edu/abs/2026PhRvL.136v0402B}
}

@article{Sommers2026Haar,
  title = {Spectral Properties and Coding Transitions of Haar-Random Quantum Codes},
  author = {Sommers, Grace M. and Jacoby, J. Alexander and Weinstein, Zack and Huse, David A. and Gopalakrishnan, Sarang},
  journal = {PRX Quantum},
  volume = {7},
  issue = {2},
  pages = {020328},
  numpages = {11},
  year = {2026},
  month = {May},
  publisher = {American Physical Society},
  doi = {10.1103/d4wh-hqcp},
  url = {https://link.aps.org/doi/10.1103/d4wh-hqcp}
}

@ARTICLE{BowenLee2025MixedTOAxiom,
       author = {{Yang}, Tai-Hsuan and {Shi}, Bowen and {Lee}, Jong Yeon},
        title = "{Topological Mixed States: Phases of Matter from Axiomatic Approaches}",
      journal = {arXiv e-prints},
         year = 2025,
        month = jun,
          eid = {arXiv:2506.04221},
        pages = {arXiv:2506.04221},
          doi = {10.48550/arXiv.2506.04221},
archivePrefix = {arXiv},
       eprint = {2506.04221},
 primaryClass = {cond-mat.str-el},
       adsurl = {https://ui.adsabs.harvard.edu/abs/2025arXiv250604221Y}
}

@article{Wang2025MixedStateTO,
  title = {Intrinsic Mixed-State Topological Order},
  author = {Wang, Zijian and Wu, Zhengzhi and Wang, Zhong},
  journal = {PRX Quantum},
  volume = {6},
  issue = {1},
  pages = {010314},
  numpages = {29},
  year = {2025},
  month = {Jan},
  publisher = {American Physical Society},
  doi = {10.1103/PRXQuantum.6.010314},
  url = {https://link.aps.org/doi/10.1103/PRXQuantum.6.010314}
}

@ARTICLE{SohalPrem2025,
       author = {{Sohal}, Ramanjit and {Prem}, Abhinav},
        title = "{Noisy Approach to Intrinsically Mixed-State Topological Order}",
      journal = {PRX Quantum},
         year = 2025,
        month = jan,
       volume = {6},
       number = {1},
          eid = {010313},
        pages = {010313},
          doi = {10.1103/PRXQuantum.6.010313},
archivePrefix = {arXiv},
       eprint = {2403.13879},
 primaryClass = {cond-mat.str-el},
       adsurl = {https://ui.adsabs.harvard.edu/abs/2025PRXQ....6a0313S}
}

@ARTICLE{EllisonCheng2025,
       author = {{Ellison}, Tyler D. and {Cheng}, Meng},
        title = "{Toward a Classification of Mixed-State Topological Orders in Two Dimensions}",
      journal = {PRX Quantum},
         year = 2025,
        month = jan,
       volume = {6},
       number = {1},
          eid = {010315},
        pages = {010315},
          doi = {10.1103/PRXQuantum.6.010315},
archivePrefix = {arXiv},
       eprint = {2405.02390},
 primaryClass = {cond-mat.str-el},
       adsurl = {https://ui.adsabs.harvard.edu/abs/2025PRXQ....6a0315E}
}

@ARTICLE{LeeJianXu2023,
       author = {{Lee}, Jong Yeon and {Jian}, Chao-Ming and {Xu}, Cenke},
        title = "{Quantum Criticality Under Decoherence or Weak Measurement}",
      journal = {PRX Quantum},
         year = 2023,
        month = aug,
       volume = {4},
       number = {3},
          eid = {030317},
        pages = {030317},
          doi = {10.1103/PRXQuantum.4.030317},
archivePrefix = {arXiv},
       eprint = {2301.05238},
 primaryClass = {cond-mat.stat-mech},
       adsurl = {https://ui.adsabs.harvard.edu/abs/2023PRXQ....4c0317L}
}

\end{document}